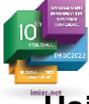

# Using Agile Story Points and Game Theory Together: Better Software Planning and Development in Agile Software Development

Bildirici Fatih[1], Codal Keziban Seçkin[2], Medeni Tunc Durmus[3]

**Abstract:** In the realm of Agile software development, precise user story point estimation is crucial for effectual project timeline and resource management. Despite its significance, the method is often marred by issues stemming from cognitive biases, disparities in individual judgment, and hurdles related to both collaboration and competition. In addressing these challenges, this study employs a comprehensive literature review, integrating key concepts from Agile software development, Story Point estimation, and Game Theory. Through rigorous examination of existing literature and relevant case studies, we identified pervasive issues in Agile and Story Point estimation. In response, we proposed the application of game theoretic strategies, notably the Vickrey Auction and Stag Hunt Game, aiming to refine these estimations. The resultant methodology not only promotes the use of game-theory inspired mechanisms but also accentuates their potential to enhance software development planning, team cohesion, and conflict resolution. Preliminary results from our research underscore the transformative potential of these games when incorporated into Agile methodologies, especially during planning and retrospective phases. The overarching goal is to achieve improved accuracy in planning, foster team collaboration, and a discernible uplift in software product quality.

# Çevik Hikaye Noktaları ve Oyun Teorisini Birlikte Kullanmak: Çevik Yazılım Geliştirmede Daha İyi Yazılım Planlama ve Geliştirme

**Özet:** Çevik yazılım geliştirme alanında, hassas kullanıcı hikayesi noktası tahmini, etkili proje zaman çizelgesi ve kaynak yönetimi için çok önemlidir. Önemine rağmen, yöntem genellikle bilişsel önyargılardan, bireysel yargılardaki eşitsizliklerden ve hem iş birliği hem de rekabetle ilgili engellerden kaynaklanan sorunlarla gölgelenmektedir. Bu çalışma, bu zorlukları ele alırken Çevik yazılım geliştirme, Hikaye Noktası tahmini ve Oyun Teorisi'nden temel kavramları entegre eden kapsamlı bir literatür taraması kullanmaktadır. Mevcut literatürün ve ilgili vaka çalışmalarının titizlikle incelenmesi yoluyla, Çevik ve Hikaye Noktası tahminindeki yaygın sorunları tespit ettik. Buna karşılık olarak, bu tahminleri iyileştirmeyi amaçlayan, özellikle Vickrey Açık Artırma ve Geyik Avı Oyunu gibi oyun teorisi stratejilerinin uygulanmasını önerdik. Ortaya çıkan metodoloji sadece oyun teorisinden esinlenen mekanizmaların kullanımını teşvik etmekle kalmıyor, aynı zamanda bunların yazılım geliştirme planlamasını, ekip uyumunu ve çatışma çözümünü geliştirme potansiyelini de vurguluyor. Araştırmamızdan elde edilen ilk sonuçlar, özellikle planlama ve geriye dönük aşamalar sırasında Çevik metodolojilere dahil edildiğinde bu oyunların dönüştürücü potansiyelinin altını çizmektedir. Genel hedef, planlamada gelişmiş doğruluk, teşvik edilmiş ekip iş birliği ve yazılım ürün kalitesinde fark edilebilir bir artış elde etmektir.

**Anahtar Kelimeler:** Çevik Yazılım Geliştirme, Oyun Teorisi, Çevik Planlama ve Tahminleme, Kullanıcı Hikayesi Noktaları

1. Department of Artificial Intelligence, Ankara University, Ankara, Turkey. Email: fbildirici@ankara.edu.tr
2. Department of Management Information Systems, Ankara Yıldırım Beyazıt University, Ankara, Turkey. Email: kseckin@aybu.edu.tr
3. Department of Management Information Systems, Ankara Yıldırım Beyazıt University, Ankara, Turkey. Email: tdmedeni@aybu.edu.tr

# INTRODUCTION

While dynamic and innovative, the multifaceted software development landscape has persistently grappled with various challenges. Among these, one of the most prominent is the difficulty in accurately estimating the time and effort needed for individual tasks, a recurring issue with potential consequences ranging from project delays to cascading effects on dependent processes (Usman et al., 2014). This persistent problem underscores the need for more refined and efficient estimation techniques in software development.

The advent of Agile methodology, a response to the rigid constraints of traditional waterfall development, offers solutions to such challenges. Agile methodologies, including Scrum, Lean, and Kanban, emphasize collaboration, customer feedback, and iterative development, instigating a radical shift in software development practices. A salient feature within this methodology, 'Story Points,' provides a relative measure of effort required to implement a feature or user story. This concept introduces a nuanced approach to handling software development's inherent complexity and uncertainty (Cohn, 2005). However, it also poses challenges, mainly when translating these estimations into tangible timelines.

Meanwhile, Game Theory, a mathematical construct designed to analyze strategic interactions among independent rational entities, is increasingly finding applications in diverse fields. Its potential application to software development, while a relatively recent exploration, offers exciting possibilities (Owen, 2013). The amalgamation of Agile Story Points and Game Theory might provide fresh insights into software development estimation processes, enabling us to navigate its complexities more efficiently.

Throughout this paper, we aim to thoroughly investigate this convergence, exploring its practical applications and potential benefits for software development. The objective is to unfold a new paradigm that promotes effective decision-making, encourages collaboration, and enhances productivity within software development practices.

# METHOD

Utilizing a literature review as our primary research method, this study integrates the principles of Agile software development, Story Point estimation, and Game Theory. We systematically examine existing literature and case studies on Agile and Story Point estimation to understand their mechanics, applications, and to pinpoint current challenges. Following this, our review extends to Game Theory, assessing its models and applications in contexts similar to software development. Drawing from this literature, we conceptualize game theoretical-inspired games, like the Stag Hunt Game and Vickrey Auction Game, to address challenges in Agile and Story Point estimation. By synthesizing insights from these diverse fields, our literature-based methodology proposes the use of game theory-derived games as potential solutions in Agile software development.

## Game Theory: Foundations and Implications for Agile Estimation

Game Theory, at its core, is a mathematical framework to understand and analyze scenarios where several players make decisions that impact not only their own outcomes but those of others. This discipline models strategic interactions and provides tools to predict outcomes based on each player's decisions and rationality. It is rooted in a set of foundational concepts:

Players: The individuals or entities making decisions within a game.

Strategies: The potential actions or decisions a player can make.

Payoffs: The outcomes or benefits received by players for each combination of strategies played.

From a structural standpoint, Game Theory isn't monolithic; it branches into various models. We see simultaneous games where decisions are made concurrently, sequential games that involve a clear order of moves, and repeated games where interaction occurs over multiple periods. Beyond the theoretical realm, the practical implications of Game Theory are profound. It's been employed across fields like economics, biology, and political science, with its essence lying in predicting rational behaviors in complex, intertwined decision-making scenarios (Spainel, 2014).

So, what might this mean for software development and, specifically, Agile story point estimation? Picture Agile estimation as a strategic 'game' – developers (or players) estimate task values (or adopt strategies) to realize optimal project deliverables (or achieve payoffs). By introducing Game Theory to this setting, biases and inaccuracies could be addressed. For instance, consider the Vickrey Auction, a non-cooperative game model. This can foster honest effort estimations by incentivizing genuine declarations (Ausubel & Milgrom, 2006).

In summary, a nuanced understanding of Game Theory's structural and practical facets can pave the way for significant enhancements in Agile software development practices, particularly during story point estimations.

## Game Theory and Usage in Software Development

Game theory is a branch of mathematics that studies the interaction of individuals' or groups' strategic decisions to predict a situation's outcome or to determine an optimal strategy. Game Theory models social situations involving conflict and cooperation between these players, assuming that they act rationally with access to complete information (Maschler et al., 2020). Their goal is to maximize their gains, which leads to a dynamic interaction of decisions that formulates the essence of the game.

As an extension of its abstract principles, Game Theory is effectively used as a decision-making tool in various situations. Its strength lies in its predictive capacity, predicting the outcomes of interactions between rational decision-makers (Siegfried, 2006). By designing hypothetical game scenarios based on real-life situations, Game Theory facilitates the derivation of optimal strategies players can adopt. Concepts such as Nash Equilibrium and Stag Hunt underpin these strategies by balancing individual and collective interests.

This theory is primarily used in economics, politics, and psychology. Game theory often involves a mathematical model of conflict and cooperation situations and is a valuable tool in various disciplines thanks to its ability to analyze decision-making processes (Samuelson, 2016). This has the potential to make it a tool that can be used in many fields beyond the relevant fields. Today, its use can be seen in many fields, from economics to biology, political sciences to sociology. In addition, the use of game theory in software development has the potential to emerge as a significant field.

In software development, game theory provides a theoretical framework for modeling and analyzing complex interactions between various stakeholders, ranging from developers and project managers to end users and attackers. Game theory can be used in various ways to improve decision-making processes and optimize resources in software engineering. Designing incentive mechanisms to ensure aligned interests among the various stakeholders in the software development process, shaping product design and marketing strategies by modeling user and competitor behavior, developing compelling test cases to detect defects, and improving resource allocation and decision-making are just some of the benefits that can be achieved with game theory (Shoham, 2008). This framework can help software engineers create more effective and efficient development processes and develop high-quality software products.

In the complex and often resource-constrained field of software development, the application of game theory provides a potent framework for strategic decision-making and resource allocation. Leveraging its mathematical approach, game theory aids in systematically managing resources, enabling the determination of task prioritization under limiting conditions (Yilmaz & O'Connor, 2010). For instance, it facilitates decision-making regarding task allocation among a finite number of engineers. Furthermore, it presents a practical approach to analyzing team dynamics and managing the intricate blend of cooperation and competition that commonly characterizes software development teams.

On a technical front, game theory finds utility in various aspects of software development. In software testing and quality assurance, it can direct focus toward components necessitating rigorous testing, providing a balanced perspective on the associated risks and rewards. Cybersecurity has witnessed the value of game theory in the design of secure protocols, the prediction of potential threats, and the strategic formulation of defense mechanisms. Similarly, requirement engineering benefits from its application in negotiating and prioritizing software requirements, considering their value and cost. Game theory also aids in prioritizing maintenance tasks based on the associated risks and potential rewards (Karabıyık et al., 2020). In the ever-evolving fields of human-computer interaction, machine learning, and artificial intelligence, game theory supports modeling interactions and predicting outcomes, thereby contributing to system designs that anticipate user needs and promote optimal interaction between AI entities (Tennenholtz, 2002). As such, incorporating game theory within the various facets of software development enhances efficiency and strategic decision-making, thereby bolstering overall project outcomes. In addition, game theory has critical application areas in Agile software development, one of the innovative software development models.

## Agile Software Development and Game Theory

Agile software development is a methodology that emphasizes flexibility, collaboration, and continuous improvement. This approach values individuals and interactions over processes and tools, focusing on working software and customer collaboration. Based on the principles of the Agile Manifesto, agile development, which responds to change rather than following a plan, encourages collaboration and communication between different stakeholders (Abrahamsson et al., 2017).

Agile methodologies are based on iterative development, where solutions evolve through the collaborative efforts of self-organizing, cross-functional teams. They encompass various techniques such as Scrum, Extreme Programming (XP), and Lean Development (Martin, 2019). By eschewing rigid planning in favor of adaptive, incremental changes, Agile Software Development facilitates more flexible responses to change and promotes customer satisfaction by continuously delivering functional software.

Agile software development  has significantly redefined project management and software engineering fields by popularising incremental and iterative processes prioritizing adaptability and customer satisfaction. Traditional development methods, often linear and rigid, have been replaced by agile methodologies that encourage flexibility, collaboration, and rapid response to change (Dingsøyr et al., 2012). Although highly effective, these methods still present numerous challenges and complexities whose understanding could benefit significantly from interdisciplinary approaches. A fascinating lens through which to view Agile software development is game theory, a mathematical field that studies strategic interactions between rational decision-makers.

Integrating game theory into the Agile Software Development framework offers an interesting analytical paradigm. Characterized by its adaptive and customer-centric methodologies, agile presents complexities that require innovative analytical tools. Game theory, a mathematical study of strategic interactions, provides a robust and insightful mechanism to explore these challenges (Gavidia-Calderon et al., 2020). This framework reveals underlying team dynamics, illuminates optimal decision-making processes, and paves the way for strategic optimization in agile methodologies.

Agile Software Development and game theory coalesce to form an insightful framework that elucidates the complex dynamics inherent in the software development lifecycle. Foremost among these dynamics is the interplay among team members, where game theory provides a robust foundation for understanding behaviors, motivations, and the alignment of individual and collective goals. This strategic viewpoint extends to critical project decisions, enabling teams to make optimal choices in task prioritization, resource allocation, and process adaptation (Lindsjørn et al., 2016). Game theory also aids in fostering more effective negotiation and conflict resolution strategies, optimizing project planning and estimation processes, and creating robust risk management systems (Tadelis, 2013).

Beyond these strategic considerations, the technical facets of Agile software development can also benefit from game theory. It contributes to designing more efficient algorithms, particularly in decision-heavy areas, and informs a more strategic allocation of testing resources to enhance software quality. Game theory also fosters a more systematic approach to decisions in software architecture, such as component coupling and cohesion. Another critical area is where game theory proves instrumental in designing incentive structures that cultivate collaboration, productivity, and quality (Hasnain et al., 2013). Furthermore, game theory helps frame the interaction between the development team and the client or stakeholders, fostering a relationship that is both cooperative and aligns with project goals.

The seamless integration of game theory into Agile software development promotes an environment of strategic optimization, enhancing project outcomes, team dynamics, and technical aspects of the software development process. The benefits of this interdisciplinary approach are far-reaching, providing avenues for improved efficacy and success within Agile software development (Stevens et al., 2021). The user story point method, a severe instrument of agile methods in planning and forecasting, is one of these areas. In this study, the framework for better use of game theory methods for software development and planning with agile story point and the benefits it will provide can be revealed more clearly.

## Better Estimation and Planning for Software Development in Collaboration with User Story Point and Game Theory

User Story, a cornerstone of Agile software development, is a succinct description of a feature from a user's viewpoint. Characteristically expressed as "As a [role], I want to [function] so that I can [benefit]," it centralizes user needs and expectations. For instance, "As an online shopper, I want to review my product choices in a basket." User Stories emphasize user-centricity, augment communication via clear and comprehensible language, and offer a flexible alternative to rigid requirement documentation (Coelho & Basu, 2012). Furthermore, they streamline planning and prioritization by disintegrating features into independent, manageable units.

A User Story Point quantifies the complexity, uncertainty, or risk associated with a User Story. It is an abstract indication of requisite work, eschewing the confines of exact temporal estimates. The utility of Story Points resides in their scalability—providing a measure of relative feature size or task complexity—thereby acknowledging the differential effort required for disparate tasks (Patton & Economy, 2014). They bolster the precision of future task duration forecasts and facilitate tracking team productivity or 'velocity' over time.

Agile methodologies, notably Scrum and Kanban, heavily utilize User Stories and Story Points. The creation of a User Story necessitates the identification of the user and their desired function, followed by the crafting of the story. Estimation of Story Points, typically a collaborative endeavor, involves team members ascribing a Story Point value to the complexity and uncertainty of a User Story, using techniques such as Planning Poker (Cohn, 2005). The aim is to harmonize varying estimates into a consensus, thereby fostering a more accurate and holistic estimate.

This estimation and planning session is done in groups; reaching a consensus is essential (Gandomani et al., 2019). In this sense, there are many problems with this approach. In this section, we will take a quick look at these problems in software development and planning of development cycles, the user story point method for better software development, and how to find solutions in the context of game theory:

**Inaccuracy in Estimates:** Estimates can often be inaccurate for various reasons, including lack of knowledge, experience, or technical understanding. Even though it is a team of experts, it cannot be easy to handle new user stories and look at this estimate of the person assigned to the task (Evbota et al., 2016). This can be addressed through repetitive games that developers learn over time to improve their predictions. This would be a game where each sprint is where team members give their predictions and are rewarded based on how accurate they are. Over time, the goal may be for team members to improve their forecasting skills.

**Groupthink:** Teams can often agree with the safest guess or dominant member, leading to biased guesses. One possible solution here where a game theory-based game would take shape would be to have anonymous guesses and then have a mechanism to discuss inconsistencies (Mayo-Wilson et al., 2010).

**Incentives:** Developers may overestimate to look good or gain more time. This is where game theory shines. We can align individual incentives with team goals by creating a game where the best strategy for the individual is to make their best guess (Easley & Ghosh, 2016). This could be a point system where the predictors are earned, not for hitting the points estimates.

**Lack of Knowledge:** Some team members may not fully understand a user story due to a lack of technical understanding, leading to incorrect predictions (Sharma & Bhattacharya, 2013). The Revelation Principle in game theory states that in well-designed games, players will find it in their best interest to reveal private information truthfully. To counter the lack of information, we can set up an information-sharing game where developers are incentivized to truthfully reveal what they do not understand about a user story and seek clarifications.

**Future Uncertainty:** Uncertainty about potential changes or obstacles can lead to inaccurate forecasts (Reneke, 2009). We can use the concept of Stochastic Games, which are games that change in a probabilistic way over time. To counteract future uncertainty, an Adaptive Forecasting Game can be designed (Giraitis et al., 2013). Here, initial predictions can be revised based on new information obtained during the development phase, rewarding adaptability. This approach also aligns with the main principles of the user story point.

**Social Loafing:** If forecasts are made collectively, some members may 'free ride' on the work of others and not try to make forecasts. This can lead to dissatisfaction, errors, unhealthy forecasts, and problems in development (Zhu & Wang, 2019). To overcome social loafing, a contribution-based reward game can be implemented. Team members are rewarded for contributing to the forecasting process, thus encouraging everyone to participate actively.

As a result, game theory offers a rich set of tools to increase the efficiency and effectiveness of user story point estimation and planning in Agile software development. Understanding and structuring incentives and rewards appropriately can reduce common challenges such as prediction inaccuracies, groupthink, fixation bias, and social loafing. In addition, game theory designs and methods in this field can offer perspectives that will solve other problems of Agile software development, such as conflict resolution during sprint retrospectives, decision-making in backlog prioritization, and collaboration strategies in distributed teams (Hasnain et al., 2013).

## Game Design Approaches to Enhance Prediction and Planning in Software Development: A Case Study

In the intricate ecosystem of software development, game theory unveils a toolbox of strategies to address pressing challenges. Predominantly, it offers frameworks for improved decision-making, effective planning, and nurturing team cohesion. Drawing from the methods detailed previously, this section introduces innovative game designs targeted at the cognitive and competition-driven hurdles faced when employing agile story points. Through these game-inspired solutions, we aim to fortify pivotal managerial aspects like forecasting and planning, which, when undermined, become precursors to project failures.

The Estimation Quagmire in Software Teams

Software development, a microcosm of broader societal structures, isn't exempt from competitive dynamics. Developers, driven by various motivations, may inflate or deflate their time estimates. This results in a domino effect of incorrect forecasts, imbalanced work distribution, spiraling technical debts, and an overall jeopardized project outcome.

> ➢ **Estimation Accuracy Game (inspired by Stag Hunt)**

Objective: Encourage accurate task estimation by developers.

Players: Developers assigned to a project.

Mechanics: Developers are tasked with estimating the time for their respective assignments. Their choices: a candid estimate or an inflated/deflated one.

Game Theory Foundation: Derived from the Stag Hunt—a model of societal decision-making where players can choose communal benefits or personal gains. Here, a genuine estimation equates to community cooperation, while skewing numbers symbolizes personal gain pursuits.

Payoff Matrix for the Estimation Accuracy Game:

| *Developer Choice* | *All Developers Cooperate* | *Some Developers Defect* |
|---|---|---|
| *Cooperate (Accurate Estimate)* | High Payoff (5 points) | Low Payoff (2 points) |
| *Defect (Inaccurate Estimate)* | Medium Payoff (3 points) | No Payoff (0 points) |

Gameplay:

Developers are tasked with estimating the duration for their assigned tasks. Once every developer has provided their estimate, the points are allocated based on:

The accuracy of their individual estimates.

The collective accuracy of all developers.

Interpretation of the Payoff Matrix:

Cooperation: If every developer provides accurate estimates (cooperates), each is rewarded with the highest payoff of 5 points. This represents the optimal scenario where all team members work in harmony, leading to precise project planning and execution.

Partial Defection: If some developers provide inaccurate estimates while others are accurate, the ones who cooperated get a lesser reward of 2 points. This serves to discourage unilateral deviations from honesty.

Unilateral Defection: If a developer defects (provides an inaccurate estimate) when every other developer cooperates, the defector receives a medium payoff of 3 points. However, this is still lesser than the payoff for mutual cooperation, thus highlighting the benefits of working in alignment with the team.

Mutual Defection: When several developers submit inaccurate estimates, those who defected are not rewarded at all, receiving 0 points. This emphasizes the detrimental consequences of collective inaccuracies in software estimation.

Benefits:

Promotion of Honesty: The game clearly rewards collective accuracy, nudging developers toward honest estimations.

Mitigation of Personal Gains: By offering a lower payoff for defecting when others cooperate, the game minimizes the individualistic tendency to inflate or deflate estimates for personal gain.

Project Health: Ensuring that estimations are accurate from the get-go translates to a more streamlined development process, minimized technical debts, and improved project timelines.

In essence, the Estimation Accuracy Game serves as a gamified tool to foster collaboration and accuracy, which are indispensable for the success of any software project.

**Problem:** Teams can often agree with the safest guess or dominant member, leading to biased guesses. One possible solution here where a game theory-based game would take shape would be to have anonymous guesses and a mechanism to discuss inconsistencies.

➢ **Vickrey Agile Estimation Game (inspired by Vickrey Auction Game)**

Objective: Counteract bias in estimations, especially those influenced by dominant team members.

Players: Development team members.

Mechanics: Developers anonymously submit their task estimates. Once sealed, these estimates are irreversible.

Game Theory Foundation: Mimicking the Vickrey Auction, the final project estimate is adopted from the second-highest bid, ensuring that extreme outliers don't skew the average.

Benefit: The VAEG ensures that no single estimation, whether astronomically high or critically low, influences the final consensus. This leads to more balanced planning and an enhanced software development process over repeated iterations.,

Payoff Matrix for the Vickrey Agile Estimation Game:

| Developer's Estimate | Actual Effort | Payoff |
|---|---|---|
| Same as Actual | Actual Effort | 5 points |
| Within +/- 10% | Actual Effort | 3 points |
| More than +/- 10% | Actual Effort | 0 points |

Gameplay and Actions:

During sprint planning, developers independently and anonymously submit their estimates for a particular user story. These estimates, once registered at a centralized repository, are irrevocable.

The crux of this game lies in the manner of selecting the final estimate. Drawing from the Vickrey Auction principles, the estimate that's chosen as the final one for the user story is the second-highest bid.

Interpretation of the Payoff Matrix:

Exact Accuracy: If a developer's estimate precisely matches the actual effort required, they are awarded the highest payoff of 5 points. This highlights the emphasis on pin-point accuracy.

Near Accuracy: Estimates that are within a 10% deviation (either way) from the actual effort grant developers a payoff of 3 points. This encourages developers to be as close to the actual estimate as possible, while still rewarding them for being in the ballpark.

Significant Deviation: For estimates that deviate by more than 10% from the actual effort, developers receive no payoff (0 points). This serves as a deterrent for wildly inaccurate guesses.

Benefits:

Reduction of Extremes: By considering the second-highest estimate, the game eliminates the influence of any extreme outliers—whether they're vastly inflated or significantly understated.

Encouragement of Genuine Estimates: The reward structure promotes accurate estimations without the pressure of conforming to a perceived average or the influence of dominant voices in the team.

Enhanced Project Planning: As developers become attuned to the game's mechanics and rewards, the collective estimations tend to converge closer to reality, resulting in improved project timelines and execution.

In a nutshell, the Vickrey Agile Estimation Game integrates auction principles with software development, creating a robust framework for teams to enhance their prediction accuracy and streamline the development process.

While our game-based strategies offer fresh insights, methods like Planning Poker, Wideband Delphi, and Fist of Five remain foundational to project estimation. Integrating these time-tested techniques with our game-theoretic designs can create a comprehensive and effective estimation system. However, the profound interplay and potential synthesis of these methodologies is a vast topic, too expansive for the scope of this paper. Truly capturing the depth and nuances warrants a dedicated exploration in a separate comprehensive study.

## FINDINGS

Integrating game theory with Agile Software Development has surfaced as a robust strategy for amplifying optimization in the software development process. Our analysis underscores the potential of game theory in guiding pivotal decision-making areas, from resource allocation and algorithm design to software architecture. Most notably, it unveils a pioneering methodology for crafting incentive systems that propel teamwork, software quality, and overall productivity. It also streamlines the often-complex interactions between the development crew and the stakeholders.

Diving into the nuances of User Story Points estimation—a cornerstone of agile practices—the inclusion of game theory has shown a marked improvement in forecasting accuracy. It addresses persistent pain points like estimation inaccuracies, the pitfalls of groupthink, and knowledge disparities within teams. Our examination further introduces innovative game-based strategies, taking cues from game theory, to refine prediction and planning in Agile settings. For instance, the 'Estimation Accuracy Game', with its roots in the 'Stag Hunt' concept, crafts a reward mechanism that prioritizes accurate task estimates. Meanwhile, the 'Vickrey Agile Estimation Game' borrows from the Vickrey Auction, ensuring unbiased estimations by choosing the penultimate estimate as the final number for a user story. These methods, backed by game theory, stand as promising antidotes to typical estimation challenges, laying the foundation for a more streamlined, transparent, and cooperative software development ecosystem.

## DISCUSSION AND CONCLUSIONS

The exploration of game theory principles within the realm of Agile software development has carved out a new frontier in enhancing story point estimation. Echoing our preliminary hypothesis, our findings emphasize that a thoughtful fusion of game theory with Agile can amplify its efficiency and output quality.

In essence, the proposed integration offers a blueprint for elevating the Agile software development process. Targeting enhanced estimation precision, fostering a collaborative team spirit, and fine-tuning project planning and delivery, the results of our exploration have been notably influential. By addressing cognitive biases, championing a cohesive team environment, and endorsing informed decision-making, this novel approach is geared towards crafting superior software solutions.

However, a word of caution—embedding this innovative approach within established Agile practices necessitates a meticulous, phased execution. It paves the path for future studies to dig deeper, refining the game theory's application in Agile and probing into its effects across diverse project genres and team dynamics. In wrapping up, our investigation resonates with the transformative potential of game theory for Agile story point estimations, suggesting a potential paradigm shift in Agile software development. While this is a burgeoning journey, our findings fortify the premise of game theory as a valuable tool in the Agile toolkit, urging further pragmatic exploration.